\documentclass[manuscript]{aastex}

\shorttitle{LiH depletion}
\shortauthors{Bovino et al.}
\begin{document}

\title{LiHe$^+$ in the early Universe: a full assessment of its reaction network and final abundances}

\author{Stefano Bovino$^1$, Roman \v{C}ur\'ik$^3$, Daniele Galli$^2$, 
Mario Tacconi$^1$, Francesco A.~Gianturco$^1$}
\affil{$^1$Department of Chemistry, the University of Rome ``Sapienza'', 
P.le A. Moro 5, 00185 Roma, Italy\\$^2$INAF-Osservatorio Astrofisico di Arcetri, Largo E. Fermi 5, 50125 Firenze, Italy\\$^3$J. Heyrovsky Institute of Physical Chemistry, Dolejskova 3, Prague, Czech Republic}

\email{fa.gianturco@caspur.it}
 
\begin{abstract}
We present the results of quantum calculations based on entirely 
ab initio methods for a variety of molecular
processes and chemical reactions involving the LiHe$^+$ ionic polar
molecule.  With the aid of these calculations we derive accurate
reaction rates and fitting expressions valid over a range 
of gas temperatures representative of 
the typical conditions of the pregalactic gas. With the help 
of a full chemical network, we then compute
the evolution of the abundance of LiHe$^+$ as function 
of redshift in the early Universe.
Finally, we compare the relative abundance of LiHe$^+$ with that of 
other polar cations formed in the same redshift interval.
\end{abstract}

\keywords{Early Universe -- Astrochemistry -- Molecular processes -- ISM: molecules --
Methods: numerical -- evolution}

\section{Introduction}

The suggestion that chemical processes involving lithium atoms, as part
of simple diatomic species formed with partner atoms and ions of expected
significant abundances (i.e. H, He, H$^+$ and He$^+$), could play a
role in the evolution of the early Universe has been put forward
several years ago \citep[e.g. see][]{319367889,319367895,319367886}. The formation of Li-bearing molecules results
from the production of Li atoms a few minutes after the Big Bang and
the concurrent formation of other light atoms like H, D and He \citep{319367898,319367891}. The fractional abundances of these
elements are sensitive to the values of the baryon density and temperature of the early
Universe and therefore imply specific physical constraints to their
actual values \citep{319367884}. In fact, as the Universe expanded
its radiation temperature decreased and the atomic ions which had
originated from the above elements gave rise to neutral atoms from ionic recombination with ambient electrons, thereby producing molecular
species chiefly by radiative association. Neutral molecules include
H$_2$, HD, and LiH.

Because of its large dipole moment and low ionization potential, LiH
has been considered a potential candidate for inducing spatial
and/or spectral distortions of the Cosmic Background Radiation (CBR),
as originally suggested by \citet{319367885} and observationally tested
by \citet{319367890} and \citet{319367892}. It is, in fact, its
possible role as a molecular coolant of the primordial gas because of
the efficient radiative decay down its manifold of rotovibrational
levels that made its likely presence a very important issue in several
earlier studies \citep{319367876}. By the same token, other
molecular species formed as cations after atomic ionization during
gravitational collapse, e.g. LiH$^+$, HeH$^+$, HD$^+$ and LiHe$^+$,
could also play a similar role as additional molecular coolants. 
Furthermore their non-equilibrium level populations may also have left a possible
signatures in protogalactic clouds, imprinting spatial or spectral
distortions in the CBR spectrum \citep{319367893}.

In particular, in this work we analyse LiHe$^+$ via a series of quantal calculations to determine the relative
role of several pathways to its formation and destruction within the
chemical network acting at low redshift.  In previous paper \citep{319367877,319367878} we examined the photonic paths to the formation and destruction
of LiHe$^+$. 
In the present paper we extend that
work here by adding the chemical paths presiding over its evolution, i.e. the chemical reaction with hydrogen:
\begin{equation}
\rm LiHe^+ + H \rightarrow LiH^+ + He
\end{equation}
leading to its destruction, and the formation reaction from the lithium
hydride cation:
\begin{equation}
\rm LiH^+ + He \rightarrow LiHe^+ + H,
\end{equation}
the physical characteristics of which are discussed extensively in our recent work
on this system \citep{319367896}.

An additional path that we shall further consider in the
present study is that driven by the presence of a residual electron fraction in the early
Universe \citep{319367895}, i.e. the fragmentation of the polar
cation by the dissociative recombination (DR) process:
\begin{equation}
\rm LiHe^+ + e^- \rightarrow LiHe^* \rightarrow Li + He + h\nu
\end{equation}
a reaction that, from the present calculations, shall turn out to be fairly efficient.

In a previous study of the lithium chemistry in the early Universe \citep{319367880}, we
found that the abundances of LiH and LiH$^+$ were of the order of about
$10^{-17}$ (number fraction), i.e. too diluted for direct detection
with presently available tools. Hence it is of interest to see what the
abundance would be for the LiHe$^+$, a molecular partner not 
considered in previous studies because of lack of reliable data. To
this goal, we shall avail ourselves of the new, and accurate, 
cross sections and reaction rates obtained here with 
fully quantal methods.

In the next section we will briefly summarize the computational
findings for reactions (1), (2), and (3) and report the corresponding
computed reaction rates at the relevant temperatures, while in section
3 we will discuss the relative role of the formation/destruction
channel rates of LiHe$^+$ at low $z$, together with its fractional
abundance compared to those of other molecules present in the early
Universe. Finally, in section 4 we shall summarize our conclusions.

\section{The computed quantum rates}

As mentioned in the introduction, our previous study on the
photon-induced evolution of LiHe$^+$ under early Universe conditions
\citep{319367877,319367878}, has shown efficient formation of the latter
molecule by resonant radiative recombination but also competitive
destruction by photodissociation at low redshifts. Hence, it was
suggested by the ensuing modeling of its relative abundances that the
LiHe$^+$ number fraction at $z\sim 30$--$1$ would be lower than 10$^{-18}$
by photon-induced paths only, i.e. much lower that the corresponding values obtained
for LiH$^+$ ($\sim 10^{-14}$) and for HeH$^+$ ($\sim 10^{-9}$) under the same conditions. On the
other hand, a more realistic picture should be obtained by the
additional implementation to the evolution scheme of the chemical routes outlined in the previous section,
in order to achieve a fuller comparison with the data which already exist for the other
polar cations formed in the early Universe.

The chemical reaction route for both processes:
\begin{equation}
\rm LiH^+(^2\Sigma^+) + He (^1S) \rightleftharpoons LiHe^+ (^1\Sigma^+) + H (^2S)
\end{equation}
of formation (with a threshold) and of destruction (without threshold)
have been studied in detail earlier by us \citep{319367896}
using accurately computed potential energy surfaces for the reactive
paths \citep{319367899}. The quantum reactive cross sections were
obtained using a Negative Imaginary Potential (NIP) approach within a
Coupled-States (CS) description of the quantum dynamics \citep{319367879}
. Full details of the computations have been reported by
\citet{319367897} and in earlier papers on ionic reactions \citep{319367881,319367879}: we therefore
refer the interested reader to those papers for further details. In
order to put the present results in a more specific context, we show in figure~1 the behaviour of the computed rates for
the formation (left panel) and destruction (right panel) reactions
given by eq.~(4) and obtained from our recent calculations \citep{319367896}.

It is interesting to note at the outset the contrasting behaviour
between the formation of the cation, which presents an energy barrier of
$\sim$ 0.05 eV \citep{319367896} and therefore becomes negligibly small
at lower redshifts, and the same process when the formed LiH$^+$ is
taken to be vibrationally excited (dashed line in the left panel).
Since the amount of internal energy is sufficient to overcome the
barrier to reaction, the formation process now shows a rather mild
dependence  on temperature and exhibits much larger formation rates
around 10$^{-10}$~cm$^3$~s$^{-1}$.

Entirely different, however, is the behaviour of the exothermic
destruction reaction, reported on the right panel of figure~1. In that
case the reaction rates are largely independent of temperature but,
even for LiHe$^+$ in its ground rotovibrational state, remain of the
order of 3-4 $\times 10^{-10}$ cm$^3$~s$^{-1}$ at the lower temperature values. It is therefore natural to expect that the different sizes of
the two reaction channels will play an important role when generating
the final abundances, as we shall discuss further in the next section.

Another ``chemical" process which we need to consider concerns the
effects of the ambient electrons on the formed cationic molecule
as described by reaction (3), i.e. the destruction of LiHe$^+$ following the
DR mechanism outlined there. The DR process has been studied before and
corresponding data under early Universe conditions already exist for
HeH$^+$ \citep{319367887} and LiH$^+$ \citep{319367883}.
No data, however, were available for the present system.

Therefore we have analysed the electron-assisted destruction of LiHe$^+$, with the ensuing
production of Li and He atoms (the details of the calculations will be presented elsewhere
\citep{new}. Here, it suffices to say that the recombinative
process, similar to the case of LiH$^+$ \citep{319367883} is
not a direct curve-crossing process with a dissociative, neutral
potential but rather an indirect process which involves Rydberg states
of the lithium atom perturbed by the presence of He. The final
destruction is therefore caused by vibrational Feshbach resonances
within the (Li$^+$He)$^*$ collisional complex which lead to the
efficient break-up between a radiatively stabilized lithium and a
neutral helium atom \citep[see][for further details]{new}.

The general  behaviour of the computed rates as a function of the temperature of the gas $T_g$ (assumed equal to the electron temperature), is
given by the calculations shown in Figure 2.  The figure clearly
indicates the marked dependence of such rates on the internal energy of
the initial target, while the definition of a global equilibrium
temperature is the one which can be used within the calculation
of the relative abundances, as we shall report in the following
section.


\section{The chemical network and the evolutionary modelling}

The evolution of the pregalactic gas is usually considered within the
framework of a Friedmann cosmological model and the cosmological
abundances of the main atomic components are taken from standard Big Bang
nucleosynthesis results \citep{319367894}. The numerical values of
the cosmological parameters used in the calculation are obtained from
WMAP5 data \citep{319367888}. For additional details on the model,
see \citet{319367886,319367882,319367880}.

In order to obtain the abundances of LiHe$^+$ which evolve from the
network of the relevant processes, a set of differential coupled rate
equations of the form:
\begin{equation}
\frac{dn_i}{dt} = \alpha_{\rm form}n_jn_k -\alpha_{\rm dest}n_i + \ldots
\label{evol}
\end{equation}
has been solved. In eq.~(5), $\alpha_{\rm form}$ and $\alpha_{\rm
dest}$ are the formation and destruction rates of the species under
discussion, and $n_i$ is the number density of the reactant species
$i$. The rate coefficients involved in the lithium chemistry network
are basically the same as those reported in \citet{319367878}, with
the exception of the reactions involving LiHe$^+$ molecule which 
now come from the present quantum calculations. Table~1 shows the 
adopted reactions and the fitting formulae of the rate coefficients.
The equations governing the temperature and redshift
evolutions have been reported in earlier papers \citep[see e.g][]{319367886,319367880} and will not be repeated here.

\section{LiHe$^+$ evolution and final abundances}

As mentioned in the previous sections, it is instructive to
model, from the present data, plus the network of rate equations outlined
before, the way in which the production of LiHe$^+$ molecules evolves,
as a function of redshift, in its interplay with the coexisting
destruction channels.

The complex evolution of all the considered species is shown in Figure~3, 
where the labels $p_i$ indicate production channels, whereas the labels $d_i$
identify the destruction channels (same notation as in Table 1). The production of LiHe$^+$ is dominated
at all redshifts by radiative association of Li$^+$ with He, both spontaneous and 
stimulated (reactions $p_1$ and $p_2$, respectively). The LiH$^+$ channel
is never effective, because of a threshold at energies corresponding
to $\sim 40$~K in the reaction of LiH$^+(v=0)$ with He (reaction $p_3$). 
The reaction of LiH$^+(v=1)$ with He (reaction $p_4$), although characterized by a much larger 
rate than its $v=0$ counterpart, is also not important because of the 
rapid decay of the $v=1$ level population of LiH$^+$ at low redhifts.
The destruction of LiHe$^+$ is due to photodissociation for $z \gtrsim 60$ 
(reaction $d_1$) and to collisions with H for $z\lesssim 60$ (reaction $d_3$),
with a significant contribution from dissociative recombination (reaction $d_2$).


The above data are considered reliable down to $z$ values of about $z =
10$ since at that stage the first stars are formed and reionization
occurs, thereby invalidating the assumptions at the basis of the
chemical network. On the whole, however, we can say that the close
competition between efficient formation paths, and nearly as efficient
destruction reactions, indicates for the present system a
limitation to its final abundances in the region of redshift of
interest.

The consequences of the interplay between the different conflicting
rates discussed above could be more clearly seen in Figure 4 showing the different abundances of various
molecular cations on a log/log scale as a function of
redshift. Most of the data for species other than LiHe$^+$ have been
presented in earlier work, while the evolution of LiHe$^+$ is obtained
from the new, accurate quantum calculations of the present study.

Due to the concurrent destruction paths, it is clear from that
comparison that, at redshift values between about
20 and 10, the fractional abundance of the LiHe$^+$ ions never increases
beyond about 10$^{-22}$ number fractions, well below those of HeH$^+$ and LiH$^+$ \citep{319367880,319367879}. While the former molecule
clearly remains the most interesting candidate for experimental
observation, LiH$^+$ is still a borderline case that provides a
challenging option while LiHe$^+$ appears still below the
sensitivity of current instrumentations.

\section{Conclusions}

In the work reported in this paper we have analysed in greater detail the
molecular processes which involve a thus far poorly studied molecular cation,
the LiHe$^+(^1\Sigma^+$), and which deal with its possible formation
and destruction in the pregalactic gas through an extensive network of
photon-induced and chemically driven processes. In the range of
redshifts of interest we have carried out quantum
calculations of chemical formation/destruction reactions which have
never been considered before from realistic computational models (see
reactions (1) to (3) of section 2).

Our results clearly show the close competition
between production and destruction pathways which turn out to have very
similar efficiencies. Hence, the estimated abundances for LiHe$^+$,
within $z$ values from about 30 and 10, do not increase as highly as
those found for LiH$^+$ and HeH$^+$.



\acknowledgements{\textbf{Acknowledgements}
We thank the CINECA and CASPUR consortia for providing us with the
necessary computational facilities and the University of Rome
``Sapienza'' for partial financial support. RC acknowledges the support of the Czech 
Ministry of Education (grants OC10046, OC09079), and the Grant Agency of the Czech Republic 
(grant P208/11/0452). We further acknowledge the support from the COST network "The chemical COSMOS" CM0805.

\bibliographystyle{apj}
\bibliography{LiHe_BIB.bib}

\begin{table*}
\tiny
\caption{Reactions considered in the present study. Fitting expression
as function of the gas ($T_g$) or radiation ($T_r$) temperatures are
accurate to better than 10\% in the range $1~{\rm K}<T_g, T_r<10^3~{\rm K}$.}
\begin{flushleft}
\begin{tabular}{lllll}
\hline
&  &  reaction   &   fitted rate (cm$^3$~s$^{-1}$ or s$^{-1}$) &  ref. \\
\hline

& $p_1$ & Li$^+$ + He $\rightarrow$ LiHe$^+$ + $h\nu$ (sp.)& $\log k=-21.87-0.204\log T_g-0.153(\log T_g)^2+0.233(\log T_g)^3-
0.144(\log T_g)^4+0.0148(\log T_g)^5$ & ($a$\/) \\
& $p_2$ & Li$^+$ + He $\rightarrow$ LiHe$^+$ + $h\nu$ (st.)& $\log (k/T_r)=-22.23-0.600\log T_g-0.086(\log T_g)^2-0.082(\log T
_g)^3+0.042(\log T_g)^4-0.0073(\log T_g)^5$ & ($a$\/) \\
& $p_3$ & LiH$^+(v=0)$ + He $\rightarrow$ LiHe$^+$ + H & $k=(5.047\times 10^{-12}-2.723\times 10^{-14}T_g+7.775\times 10^{-17}
T_g^2-1.954\times 10^{-20}T_g^3)\exp(-441/T_g)$ & ($b$\/) \\
& $p_4$ & LiH$^+(v=1)$ + He $\rightarrow$ LiHe$^+$ + H & $\log k=-9.96-0.648\log T_g-2.743(\log T_g)^2+6.092(\log T_g)^3-9.297
(\log T_g)^4+8.397(\log T_g)^5$ & \\
&       &                                              & $-4.420(\log T_g)^6+1.328(\log T_g)^7-0.210(\log T_g)^8+0.013(\log T_
g)^9$ & ($b$\/) \\
& $d_1$ & LiHe$^+$ + $h\nu$ $\rightarrow$ Li$^+$ + He & $k=21\, T_r^{0.25}\exp(-5460/T_r)$  & ($a$\/) \\
& $d_2$ & LiHe$^+$ + e $\rightarrow$ Li + He & $\log k=-5.254-0.452\log T_g-0.791(\log T_g)^2+0.463(\log T_g)^3-0.076(\log T_g
)^4$ & ($c$\/) \\
& $d_3$ & LiHe$^+$ + H $\rightarrow$ LiH$^+$ + He & $\log k=5.780-2.274(\log T_g)+0.154(\log T_g)^2+0.0639(\log T_g)^3$ & ($a$
\/) \\
\hline
\end{tabular}

\vspace{1em}
($a$\/)~Bovino, Tacconi, Gianturco ~(2012); ($b$\/)~Tacconi, Bovino,
Gianturco ~(2012); ($c$\/)~\v{C}ur\'ik et al.~(2012, in preparation).
\end{flushleft}
\end{table*}

\begin{figure}
\includegraphics[width=1\textwidth]{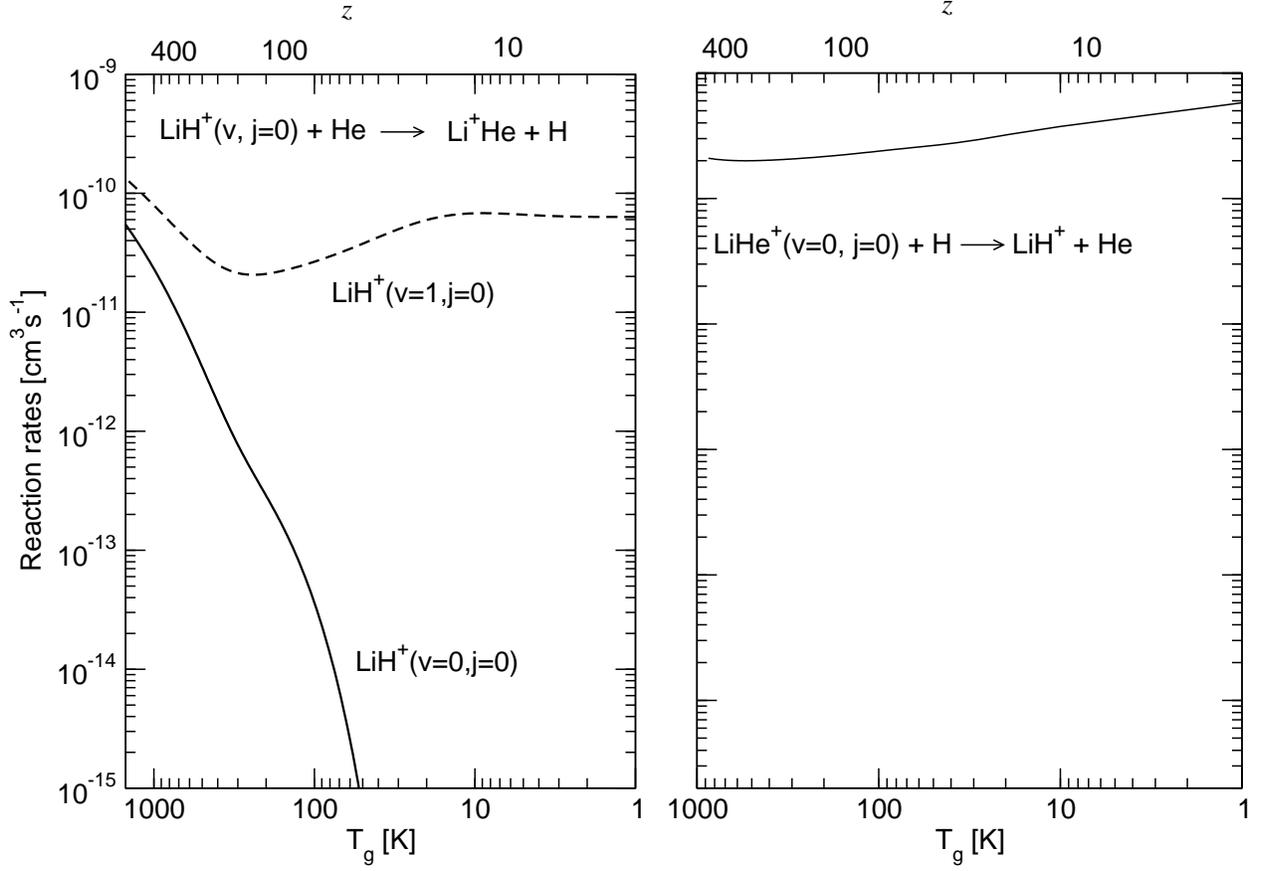} 
\caption{Computed quantum rates (from Tacconi et al.~2011a) of the
LiHe$^+$ formation reaction ({\em left panel}\/) and of its destruction by
hydrogen ({\em right panel}\/) down to the low redshifts of early Universe
environment.}
\end{figure} 

\begin{figure}
\includegraphics[width=1\textwidth]{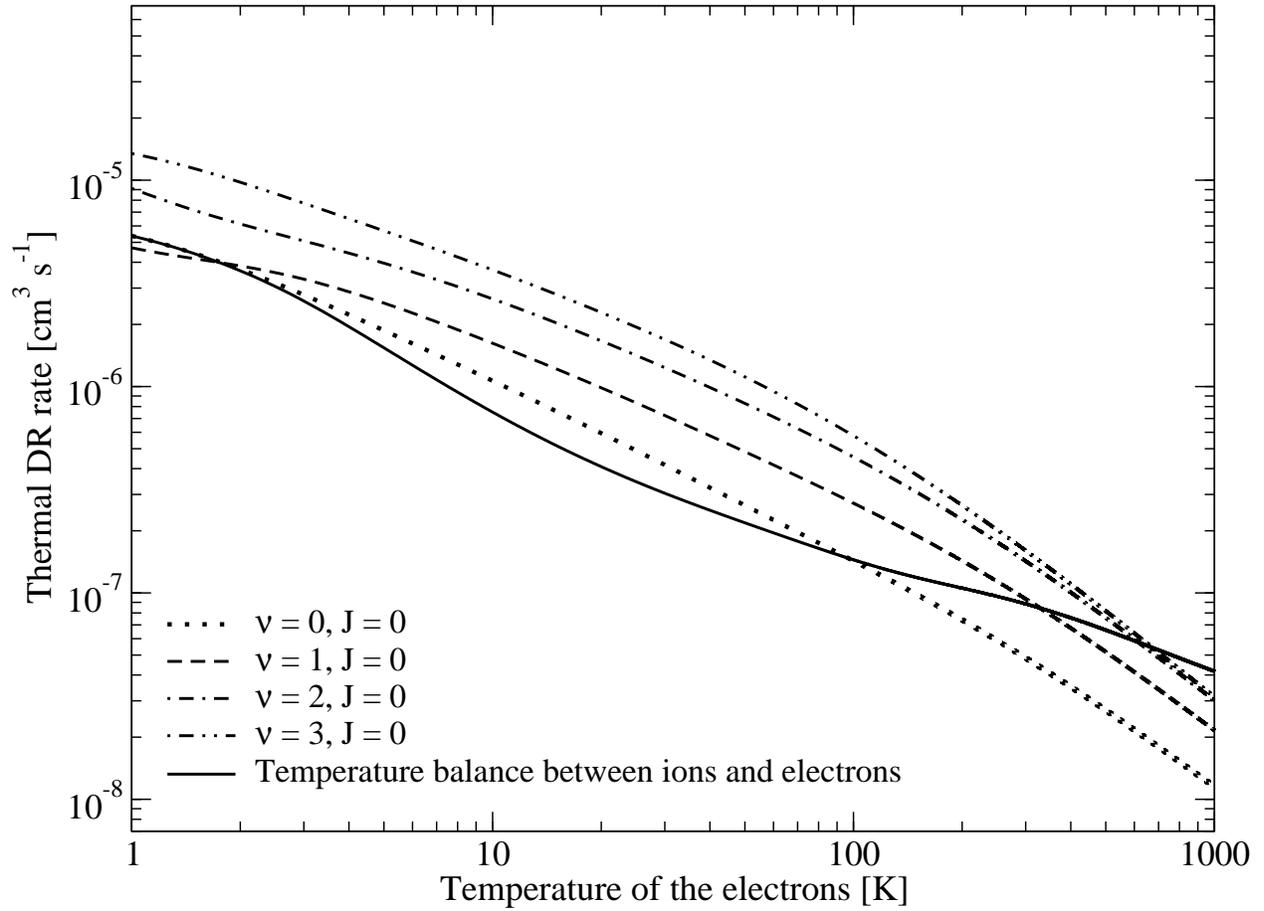}
\caption{Computed dissociative recombination (DR) rates for LiHe$^+$ as
a function of the gas (electron) temperature. Different vibrational
levels of initial LiHe$^+$ are shown, while the solid black curve
represents the case where a temperature balance exists between the
internal energy of the ion and that of the electrons.}
\end{figure}

\begin{figure}
\includegraphics[width=1\textwidth]{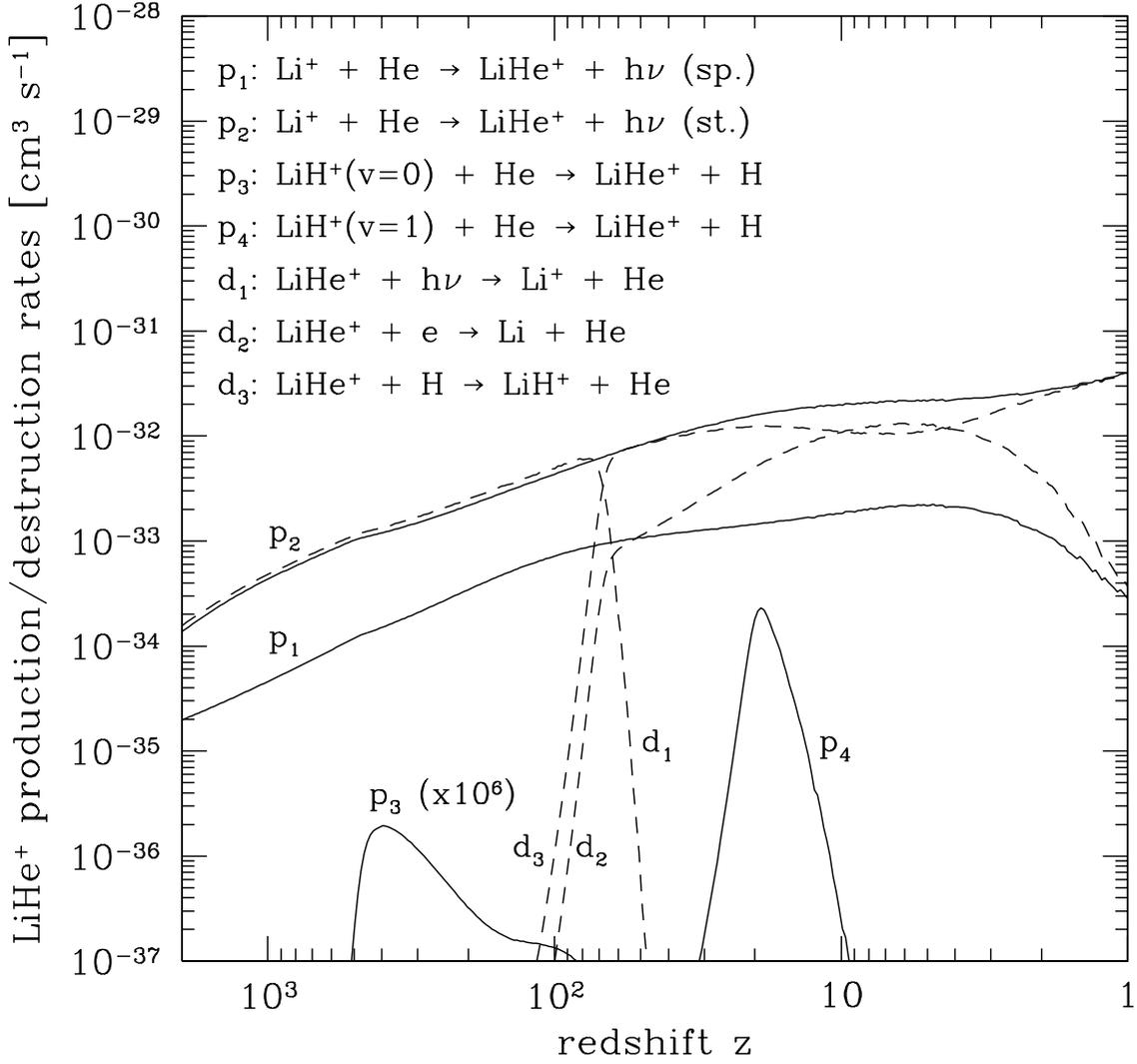}
\caption{Computed production/destruction channel evolutions for
LiHe$^+$ as a function of the redshift $z$. The present data for LiHe$^+$ include the corrections to the results by Bovino et al. (2011a) presented by Bovino et al. (2012).}
\end{figure}
   
\begin{figure}
\includegraphics[width=1\textwidth]{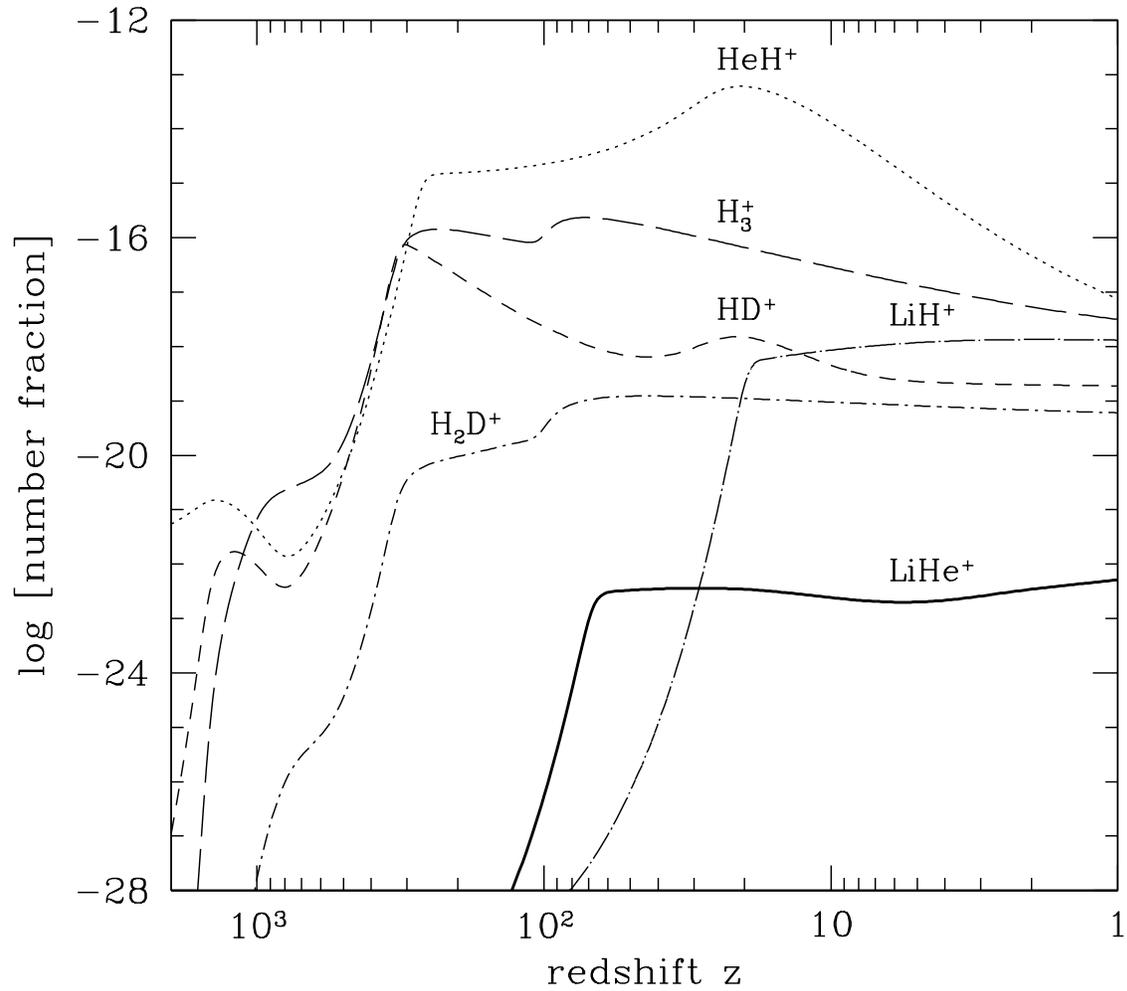}
\caption{Computed abundances (number fractions) of 
molecular cations in the early Universe as a function 
of the redshift $z$, obtained from the rates reported in Figure 3.}
\end{figure}

\end{document}